Inelastic soliton processes generated by the perturbed KdV equation

Yair Zarmi
Jacob Blaustein Institutes for Desert Research
And Physics Department
Ben-Gurion University of the Negev
Midreshet Ben-Gurion, 84990
Israel

Abstract

Perturbations commonly added to the KdV equation contain terms that represent inelastic interactions among KdV solitons in multiple-soliton solutions. These terms trigger the emergence of new waves in the first-order correction to the solution of the perturbed equation. The new waves can represent soliton-anti-soliton creation, annihilation and scattering. Other possible waves represent soliton decay and production.

The KdV equation provides a lowest-order approximation for the evolution of small-amplitude solutions of complex dynamical systems, such as the classical shallow-water problem [1], the Fermi-Pasta-Ulam problem [2] and the ion acoustic wave equations in Plasma Physics [3, 4]. When the terms neglected in the derivation of the equation are re-instated, the resulting perturbed equation often has the following generic form [5-7], given here through first order only:

$$w_t = 6 w w_1 + w_3 + \varepsilon \left( 30 \alpha_1 w^2 w_1 + 10 \alpha_2 w w_3 + 20 \alpha_3 w_1 w_2 + \alpha_4 w_5 \right) + O(\varepsilon^2) \quad (|\varepsilon| \ll 1, \ w_p \equiv \partial_x^p w) \tag{1}$$

The Normal Form perturbative analysis of the solution of Eq. (1) [5-7] entails the expansion of the solution in an asymptotic series in powers of $\varepsilon$:

$$w(t,x) = u(t,x) + \varepsilon u^{(1)}(t,x) + O(\varepsilon^2) \ , \tag{2}$$

and the assumption that the $O(\varepsilon^0)$ term, $u(t,x)$, is governed by an integrable Normal Form:

$$u_t = 6 u u_1 + u_3 + \varepsilon \alpha_4 \left( 30 u^2 u_1 + 10 u u_3 + 20 u_1 u_2 + u_5 \right) + O(\varepsilon^2) \ . \tag{3}$$

Eq. (3) has the same (single, multiple) soliton solutions as the KdV equation, with a simple modification of the velocity of each soliton. Denoting the velocity and wave number of the $i^{\text{th}}$ soliton by $v_i$ and $k_i$, respectively, $v_i$ is given by

$$v_i = 4 k_i^2 \left( 1 + \varepsilon \alpha_4 \, 4 k_i^2 + O(\varepsilon^2) \right) \ . \tag{4}$$

The multiple-soliton solution describes an elastic-collision process. Away from the origin, each soliton has the same functional form (including wave number and velocity) before and after the collision. It is unaffected by the existence of the other solitons, except for a possible trivial phase shift, which depends on the wave numbers of the other solitons [8-12].

The solution of Eq. (1) can be written as a sum of an elastic component and a first-order inelastic component [13]. The elastic component has the structure of Eq. (2), with higher-order terms that are known differential polynomials in u(t,x). It is the full solution of Eq. (1) when u(t,x) is a single-soliton solution of the Normal Form, Eq. (3). When u(t,x) is a multiple-soliton solution, an inelastic component must be included in the first-order correction in Eq. (2):

$$u^{(1)} = u_{el}^{(1)}(t,x) + \eta(t,x) \ . \tag{5}$$

The first-order contribution to the elastic component, $u_{el}^{(1)}(t,x)$, is

$$u_{el}^{(1)} = \left(-\tfrac{5}{2}\alpha_1 + \tfrac{10}{3}\alpha_2 + \tfrac{5}{3}\alpha_3 - \tfrac{5}{2}\alpha_4\right)u_2 + \left(-5\alpha_1 + 5\alpha_2\right)u^2 \ . \tag{6}$$

Being a differential polynomial in u, it does not spoil the elastic scattering picture exhibited by u when the latter is a multiple-soliton solution.

To lowest order in ε, the inelastic component, η(t,x), obeys the following equation [13]:

$$\eta_t = 6\partial_x(u\eta) + \eta_{xxx} + 10(\alpha_2 - \alpha_4)R_1[u] \ . \tag{7}$$

The driving term in Eq. (7) is the part of the first-order perturbation in Eq. (1), which is responsible for inelastic soliton interactions. In the usual version of the Normal Form analysis, it is identified as an obstacle to asymptotic integrability [5-7, 14]. $R_1[u]$ is given by

$$R_1[u] = \partial_x R_0[u] \ , \quad \left(R_0[u] = u^3 - u_1^2 + u u_2\right) \ . \tag{8}$$

Both $R_1[u]$ and $R_0[u]$ vanish identically when u(t,x) is a single-soliton solution. Consequently, by construction, they vanish exponentially fast in all directions in the x-t plane when u(t,x) is a multiple-soliton solution [14]. The inelastic nature of the effect of $R_1[u]$ on the solution shows up by the fact that $R_1[u]$ depends on the wave numbers of all solitons, so that far away from the origin, each soliton is modified by the existence of the other solitons in a non-trivial manner. For example, in the case of the two-soliton solution given by [15]

$$u(t,x) = 2\partial_x^2 \ln f(t,x)$$
$$f(t,x) = 1 + g(t,x;k_1) + g(t,x;k_2) + \left(\frac{k_1 - k_2}{k_1 + k_2}\right)^2 g(t,x;k_1) g(t,x;k_2) \ , \qquad (9)$$
$$g(t,x;k) = \exp\{2k(x + 4k^2 t)\}$$

$R_1[u]$ is proportional to $(k_1 - k_2)^2$ and has additional non-trivial $k_1$ and $k_2$ dependence.

As $R_1[u]$ is a complete differential, and generates bounded solutions for $\eta(t,x)$ [13], it is simpler to study another function, $\omega(t,x)$, related to $\eta(t,x)$ by

$$\eta(t,x) = \partial_x \omega(t,x) \ . \qquad (10)$$

Also, as Eq. (7) is linear in $\eta(t,x)$, one can replace the numerical coefficient that multiplies $R_1[u]$ in that equation by 1. The resulting equation for $\omega(t,x)$ is

$$\omega_t = 6u\omega_x + \omega_{xxx} + R_0[u] \ . \qquad (11)$$

The homogeneous equation corresponding to Eq. (11) is

$$\omega_t = 6u\omega_x + \omega_{xxx} \ . \qquad (12)$$

A possible solution of Eq. (12) is

$$\omega(t,x) = au(t,x) \ . \qquad (13)$$

Any value of $a$ is allowed, because $u(t,x)$ obeys the KdV equation at this level of approximation.

The behavior of the system under discussion far from the origin offers a new possibility for solutions of Eq. (11). First, Eq. (12) is an approximation for Eq. (11) far away from the origin, where $R_0[u]$ vanishes exponentially. Second, the multiple-soliton solution is well approximated by a sum of well-separated single solitons. For instance, in the two-soliton case,

$$u(t,x) \underset{t \to \infty}{\to} u_s(t,x;k_1) + u_s(t,x;k_2) \quad , \quad \left( u_s(t,x;k_i) = \frac{2 k_i^2}{\left(\cosh\{k_i(x + 4 k_i^2 t + \xi_i)\}\right)^2} \right) . \tag{14}$$

(The phase shifts, $\xi_i$, are known functions of $k_1$ and $k_2$.)

The new possibility is that, away from the origin, the solution of Eq. (11), asymptotes into a sum of single KdV solitons, each multiplied by a different coefficient. For example, in the two-soliton case, denoting the single-soliton solution of the KdV equation by $u_s(t,x;k)$, one may expect

$$\omega(t,x) \underset{t \to \infty}{\to} a_1 u_s(t,x;k_1) + a_2 u_s(t,x;k_2) . \tag{15}$$

The coefficients $a_2$ and $a_2$ are to be determined by the influence of $R_0(t,x)$. From the intuitive point of view, once away from the origin, so that the effect of $R_0[u]$ is not felt anymore, the only possibility for $\omega(t,x)$ to survive in Eq. (11) is if it overlaps with the soliton trajectories. A detailed numerical study, on which this paper is based, indicates that Eq. (15) seems to emerge naturally.

The examples reported in the following have been obtained for the case when $u(t,x)$ is a two-soliton solution, with $k_1 = 0.2$ and $k_2 = 0.1$. Fig. 1 shows a soliton-anti-soliton creation solution of Eq. (11), obtained for vanishing initial data at a large negative value of $t$. A study of the numerical data reveals that, for $t \gg 0$, the creation solution behaves as a sum of a soliton and an anti-soliton, which follow the exact trajectories, and have the same functional form (including wave numbers, velocities and phase shifts) of the KdV solitons. Their amplitudes differ from those the of KdV solitons, owing to the influence of $R_0(t,x)$. Fig. 2 shows the $x$-dependence of the two-KdV-soliton solution given by Eq. (9), along with the creation solution at times sufficiently long for the asymptotic behavior to prevail.

Because of the fact that of $u(t,x)$, $\omega(t,x)$ and $R_0[u]$ vanish as $t \to \infty$, at least asymptotically, the total mass of $\omega(t,x)$ is constant:

$$\frac{d}{dt}m(t) = \frac{d}{dt}\int_{-\infty}^{+\infty}\omega(t,x)dx \underset{t \to \infty}{\to} 0 \ . \tag{16}$$

Within the numerical accuracy of the computations, the soliton-anti-soliton wave created from vanishing initial data has a vanishing total mass. To the extent that this is a precise statement, Eqs. (14)-(16) imply that $a_1$ and $a_2$ ought to obey the relation

$$k_1 a_1 + k_2 a_2 = 0 \ . \tag{17}$$

Eq. (17) is obeyed by the coefficients (deduced from the numerical extrema of the soliton anti-soliton solution) within the numerical precision of the computation.

A solution that corresponds to soliton-anti-soliton annihilation is readily obtained by imposing the following initial data:

$$\omega(t,x) \underset{t \to -\infty}{\to} -a_1 u_s(t,x;k_1) - a_2 u_s(t,x;k_2) \ . \tag{18}$$

$a_1$ and $a_2$ have the values found previously in the case of the creation wave, the asymptotic form of which is given in Eq. (15). An example of soliton-anti-soliton annihilation is shown in Fig. 3.

As Eq. (11) is linear, any linear combination of the creation and annihilation solutions, denoted by $\omega_c(t,x)$ and $\omega_a(t,x)$, respectively,

$$\omega(t,x) = \mu_c \omega_c(t,x) + \mu_a \omega_a(t,x) \quad , \tag{19}$$

is a solution of Eq. (11), provided

$$\mu_c + \mu_a = 1 \quad . \tag{20}$$

Fig. 4 shows such a solution for the case $\mu_c = \mu_a = 1/2$. The result is a soliton-anti-soliton scattering process. The collision is kinemtically elastic, but involves sign exchange. Other values of $\mu_c$ and $\mu_a$ that obey Eq. (20) correspond to mass exchange collisions.

Finally, adding the solution, Eq. (13), of the homogeneous equation, Eq. (12), with an appropriate choice of the coefficient $a$, to the creation and/or annihilation solutions, leads to new solutions, such as soliton decay and binding. For instance, choosing $a = -a_2$ (with $a_2$ of Eq. (15) obtained from the numerical solution of Eq. (11) for the pair creation process, $\omega_c(t,x)$), the following is a solution that represents the generation of a "bound state" in the collision of two KdV solitons, the amplitudes of which are modified according to

$$\omega(t,x) = -a_2 u(t,x) + \omega_c(t,x) \quad . \tag{21}$$

(For $t \to +\infty$, soliton no. 2 in $u(t,x)$ exactly cancels the anti-soliton in $\omega_c$.) An example is shown in Fig. 5. In a similar manner, combining Eq. (13) with the annihilation solution, $\omega_a(t,x)$ according to

$$\omega(t,x) = a_2 u(t,x) + \omega_a(t,x) \quad , \tag{22}$$

describes a process, in which a KdV-like soliton decays into two solitons.

The physical picture delineated above is not confined to the two-soliton case. Solutions of the same nature have been obtained for the multiple-soliton case. It is easy to construct, for example, a process in which *M* incoming solitons are converted into *N* outgoing ones. All have the standard KdV structure, albeit with modified amplitudes. In addition, the effect of higher-order perturbations, through $O(\varepsilon^3)$, added to Eq. (1) has been analyzed, and does not alter this picture. The detailed results will be reported elsewhere.

The solution of Eq. (11) is determined by the initial data, for which one wishes to solve Eq. (1). The important conclusion is that, although $\eta(t,x)$ in Eq. (5) does lead to an inelastic effect (each soliton is modified by a first-order correction that depends on the wave numbers of other solitons in a non-trivial manner), it *does not* change the multiple-soliton nature of the solution. The reason is that the new waves propagate along the characteristic lines of the KdV solitons, and have the same profiles as the KdV solitons. This conclusion is in agreement with the findings of [16] on small-amplitude solutions of the ion acoustic wave equation of Plasma Physics, that the numerical solution of the full system of equations preserves the soliton nature of the KdV solutions, except for a modification of soliton velocities by the effect of the perturbation.

Acknowledgments: Many discussions with G. Burde and B. Zaltzman are gratefully acknowledged.

FIGURE CAPTIONS

Fig. 1 Soliton-anti-soliton creation solution of Eq. (11) in two-soliton case.

Fig. 2 Two-soliton solution, Eq. (9) (full line) & soliton anti-soliton creation solution of Eq. (11) (dashed line).  Thin lines - $t = 400$, Thick lines - $t = 800$.

Fig. 3 Soliton-anti-soliton annihilation solution of Eq. (11) in two-soliton case.

Fig. 4 Soliton-anti-soliton scattering solution of Eq. (11) in two-soliton case.

Fig. 5 "Bound state" production in a collision of KdV- like solitons (Eq. (21)) in two-soliton case.

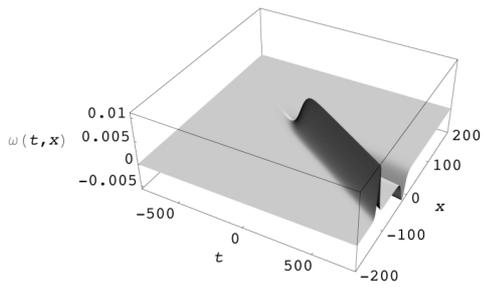

Fig. 1

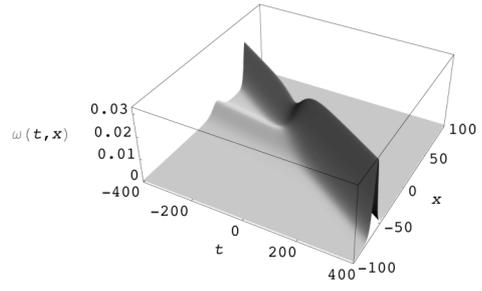

Fig.5

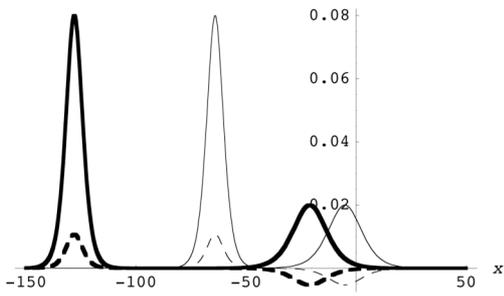

Fig. 2

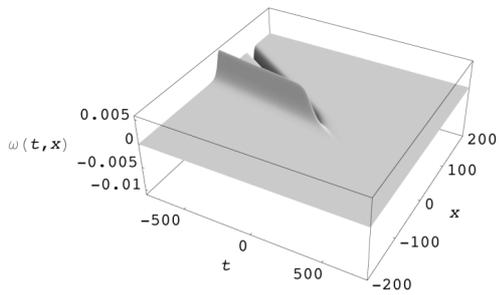

Fig. 3

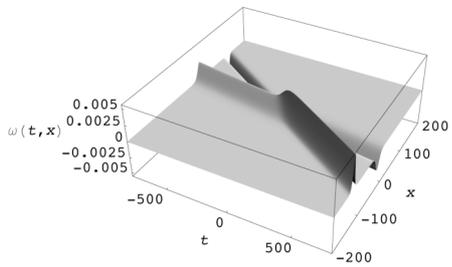

Fig. 4